\documentclass[twocolumn,showpacs,preprintnumbers,amsmath,prl,amssymb,superscriptaddress]{revtex4}

\pdfoutput=1

\usepackage{graphicx}
\usepackage{dcolumn}
\usepackage{bm}

\def\cpkkd{\rm{kg^{-1}keV^{-1}day^{-1}}}
\def\mwimp{\rm{m_{\chi}}}
\def\csnospin{\rm{\sigma_{\chi N}^{SI}}}
\def\effbs{\varepsilon _{\rm BS}}
\def\lmbdbs{\lambda _{\rm BS}}
\def\ppcge{{\it p}{\rm PCGe}}

\begin{document}


\title{Limits on light WIMPs from the CDEX-1 experiment with a p-type point-contact germanium detector at the China Jinping Underground Laboratory

}


\affiliation{Key Laboratory of Particle and Radiation Imaging (Ministry of Education) and Department of Engineering Physics, Tsinghua University, Beijing 100084}
\affiliation{College of Physical Science and Technology, Sichuan University, Chengdu 610064}
\affiliation{Department of Nuclear Physics, China Institute of Atomic Energy, Beijing 102413}
\affiliation{School of Physics, Nankai University, Tianjin 300071}
\affiliation{NUCTECH Company, Beijing 10084}
\affiliation{YaLong River Hydropower Development Company, Chengdu 610051}
\affiliation{Institute of Physics, Academia Sinica, Taipei 11529}
\affiliation{Department of Physics, Banaras Hindu University, Varanasi 221005}
\author{Q. Yue} \altaffiliation [Corresponding author: ]{yueq@mail.tsinghua.edu.cn}
\affiliation{Key Laboratory of Particle and Radiation Imaging (Ministry of Education) and Department of Engineering Physics, Tsinghua University, Beijing 100084}
\author{W. Zhao}
\altaffiliation[ Corresponding author: ]{w-zhao11@mail.tsinghua.edu.cn}
\affiliation{Key Laboratory of Particle and Radiation Imaging (Ministry of Education) and Department of Engineering Physics, Tsinghua University, Beijing 100084}
\author{K.J.~Kang}
\affiliation{Key Laboratory of Particle and Radiation Imaging (Ministry of Education) and Department of Engineering Physics, Tsinghua University, Beijing 100084}
\author{J.P.~Cheng}
\affiliation{Key Laboratory of Particle and Radiation Imaging (Ministry of Education) and Department of Engineering Physics, Tsinghua University, Beijing 100084}
\author{Y.J.~Li}
\affiliation{Key Laboratory of Particle and Radiation Imaging (Ministry of Education) and Department of Engineering Physics, Tsinghua University, Beijing 100084}
\author{S.T.~Lin}
\altaffiliation{Corresponding author: linst@phys.sinica.edu.tw}
\affiliation{College of Physical Science and Technology, Sichuan University, Chengdu 610064}
\affiliation{Institute of Physics, Academia Sinica, Taipei 11529}
\author{J.P.~Chang}
\affiliation{NUCTECH Company, Beijing 10084}
\author{N.~Chen}
\affiliation{Key Laboratory of Particle and Radiation Imaging (Ministry of Education) and Department of Engineering Physics, Tsinghua University, Beijing 100084}
\author{Q.H.~Chen}
\affiliation{Key Laboratory of Particle and Radiation Imaging (Ministry of Education) and Department of Engineering Physics, Tsinghua University, Beijing 100084}
\author{Y.H.~Chen}
\affiliation{YaLong River Hydropower Development Company, Chengdu 610051}
\author{Y.C.~Chuang}
\altaffiliation{Participating as a member of TEXONO Collaboration}
\affiliation{Institute of Physics, Academia Sinica, Taipei 11529}
\author{Z.~Deng}
\affiliation{Key Laboratory of Particle and Radiation Imaging (Ministry of Education) and Department of Engineering Physics, Tsinghua University, Beijing 100084}
\author{Q.~Du}
\affiliation{School of Physics, Nankai University, Tianjin 300071}
\author{H.~Gong}
\affiliation{Key Laboratory of Particle and Radiation Imaging (Ministry of Education) and Department of Engineering Physics, Tsinghua University, Beijing 100084}
\author{X.Q.~Hao}
\affiliation{Key Laboratory of Particle and Radiation Imaging (Ministry of Education) and Department of Engineering Physics, Tsinghua University, Beijing 100084}
\author{H.J.~He}
\affiliation{Key Laboratory of Particle and Radiation Imaging (Ministry of Education) and Department of Engineering Physics, Tsinghua University, Beijing 100084}
\author{Q.J.~He}
\affiliation{Key Laboratory of Particle and Radiation Imaging (Ministry of Education) and Department of Engineering Physics, Tsinghua University, Beijing 100084}
\author{H.X.~Huang}
\affiliation{Department of Nuclear Physics, China Institute of Atomic Energy, Beijing 102413}
\author{T.R.~Huang}
\altaffiliation{Participating as a member of TEXONO Collaboration}
\affiliation{Institute of Physics, Academia Sinica, Taipei 11529}
\author{H.~Jiang}
\affiliation{Key Laboratory of Particle and Radiation Imaging (Ministry of Education) and Department of Engineering Physics, Tsinghua University, Beijing 100084}
\author{H.B.~Li}
\altaffiliation{Participating as a member of TEXONO Collaboration}
\affiliation{Institute of Physics, Academia Sinica, Taipei 11529}
\author{J.M.~Li}
\affiliation{Key Laboratory of Particle and Radiation Imaging (Ministry of Education) and Department of Engineering Physics, Tsinghua University, Beijing 100084}
\author{J.~Li}
\affiliation{Key Laboratory of Particle and Radiation Imaging (Ministry of Education) and Department of Engineering Physics, Tsinghua University, Beijing 100084}
\author{J.~Li}
\affiliation{NUCTECH Company, Beijing 10084}
\author{X.~Li}
\affiliation{Department of Nuclear Physics, China Institute of Atomic Energy, Beijing 102413}
\author{X.Y.~Li}
\affiliation{School of Physics, Nankai University, Tianjin 300071}
\author{Y.L.~Li}
\affiliation{Key Laboratory of Particle and Radiation Imaging (Ministry of Education) and Department of Engineering Physics, Tsinghua University, Beijing 100084}
\author{H.Y.~Liao}
\altaffiliation{Participating as a member of TEXONO Collaboration}
\affiliation{Institute of Physics, Academia Sinica, Taipei 11529}
\author{F.K.~Lin}
\altaffiliation{Participating as a member of TEXONO Collaboration}
\affiliation{Institute of Physics, Academia Sinica, Taipei 11529}
\author{S.K.~Liu}
\affiliation{College of Physical Science and Technology, Sichuan University, Chengdu 610064}
\author{L.C.~L\"{u}}
\affiliation{Key Laboratory of Particle and Radiation Imaging (Ministry of Education) and Department of Engineering Physics, Tsinghua University, Beijing 100084}
\author{H. Ma}
\affiliation{Key Laboratory of Particle and Radiation Imaging (Ministry of Education) and Department of Engineering Physics, Tsinghua University, Beijing 100084}
\author{S.J.~Mao}
\affiliation{NUCTECH Company, Beijing 10084}
\author{J.Q.~Qin}
\affiliation{Key Laboratory of Particle and Radiation Imaging (Ministry of Education) and Department of Engineering Physics, Tsinghua University, Beijing 100084}
\author{J.~Ren}
\affiliation{Department of Nuclear Physics, China Institute of Atomic Energy, Beijing 102413}
\author{J.~Ren}
\affiliation{Key Laboratory of Particle and Radiation Imaging (Ministry of Education) and Department of Engineering Physics, Tsinghua University, Beijing 100084}
\author{X.C.~Ruan}
\affiliation{Department of Nuclear Physics, China Institute of Atomic Energy, Beijing 102413}
\author{M.B.~Shen}
\affiliation{YaLong River Hydropower Development Company, Chengdu 610051}
\author{L.~Singh}
\altaffiliation{Participating as a member of TEXONO Collaboration}
\affiliation{Institute of Physics, Academia Sinica, Taipei 11529}
\affiliation{Department of Physics, Banaras Hindu University, Varanasi 221005}
\author{M.K.~Singh}
\altaffiliation{Participating as a member of TEXONO Collaboration}
\affiliation{Institute of Physics, Academia Sinica, Taipei 11529}
\affiliation{Department of Physics, Banaras Hindu University, Varanasi 221005}
\author{A.K.~Soma}
\altaffiliation{Participating as a member of TEXONO Collaboration}
\affiliation{Institute of Physics, Academia Sinica, Taipei 11529}
\author{J.~Su}
\affiliation{Key Laboratory of Particle and Radiation Imaging (Ministry of Education) and Department of Engineering Physics, Tsinghua University, Beijing 100084}
\author{C.J.~Tang}
\affiliation{College of Physical Science and Technology, Sichuan University, Chengdu 610064}
\author{C.H.~Tseng}
\altaffiliation{Participating as a member of TEXONO Collaboration}
\affiliation{Institute of Physics, Academia Sinica, Taipei 11529}
\author{J.M.~Wang}
\affiliation{YaLong River Hydropower Development Company, Chengdu 610051}
\author{L.~Wang}
\affiliation{Key Laboratory of Particle and Radiation Imaging (Ministry of Education) and Department of Engineering Physics, Tsinghua University, Beijing 100084}
\author{Q.~Wang}
\affiliation{Key Laboratory of Particle and Radiation Imaging (Ministry of Education) and Department of Engineering Physics, Tsinghua University, Beijing 100084}
\author{H.T.~Wong}
\altaffiliation{Participating as a member of TEXONO Collaboration}
\affiliation{Institute of Physics, Academia Sinica, Taipei 11529}
\author{S.Y.~Wu}
\affiliation{YaLong River Hydropower Development Company, Chengdu 610051}
\author{Y.C.~Wu}
\affiliation{Key Laboratory of Particle and Radiation Imaging (Ministry of Education) and Department of Engineering Physics, Tsinghua University, Beijing 100084}
\author{Y.C.~Wu}
\affiliation{NUCTECH Company, Beijing 10084}
\author{Z.Z.~Xianyu}
\affiliation{Key Laboratory of Particle and Radiation Imaging (Ministry of Education) and Department of Engineering Physics, Tsinghua University, Beijing 100084}
\author{R.Q.~Xiao}
\affiliation{Key Laboratory of Particle and Radiation Imaging (Ministry of Education) and Department of Engineering Physics, Tsinghua University, Beijing 100084}
\author{H.Y.~Xing}
\affiliation{College of Physical Science and Technology, Sichuan University, Chengdu 610064}
\author{F.Z.~Xu}
\affiliation{Key Laboratory of Particle and Radiation Imaging (Ministry of Education) and Department of Engineering Physics, Tsinghua University, Beijing 100084}
\author{Y.~Xu}
\affiliation{School of Physics, Nankai University, Tianjin 300071}
\author{X.J.~Xu}
\affiliation{Key Laboratory of Particle and Radiation Imaging (Ministry of Education) and Department of Engineering Physics, Tsinghua University, Beijing 100084}
\author{T.~Xue}
\affiliation{Key Laboratory of Particle and Radiation Imaging (Ministry of Education) and Department of Engineering Physics, Tsinghua University, Beijing 100084}
\author{L.T.~Yang}
\affiliation{Key Laboratory of Particle and Radiation Imaging (Ministry of Education) and Department of Engineering Physics, Tsinghua University, Beijing 100084}
\author{S.W.~Yang}
\altaffiliation{Participating as a member of TEXONO Collaboration}
\affiliation{Institute of Physics, Academia Sinica, Taipei 11529}
\author{N.~Yi}
\affiliation{Key Laboratory of Particle and Radiation Imaging (Ministry of Education) and Department of Engineering Physics, Tsinghua University, Beijing 100084}
\author{C.X.~Yu}
\affiliation{School of Physics, Nankai University, Tianjin 300071}
\author{H.~Yu}
\affiliation{Key Laboratory of Particle and Radiation Imaging (Ministry of Education) and Department of Engineering Physics, Tsinghua University, Beijing 100084}
\author{X.Z.~Yu}
\affiliation{College of Physical Science and Technology, Sichuan University, Chengdu 610064}
\author{X.H.~Zeng}
\affiliation{YaLong River Hydropower Development Company, Chengdu 610051}
\author{Z.~Zeng}
\affiliation{Key Laboratory of Particle and Radiation Imaging (Ministry of Education) and Department of Engineering Physics, Tsinghua University, Beijing 100084}
\author{L.~Zhang}
\affiliation{NUCTECH Company, Beijing 10084}
\author{Y.H.~Zhang}
\affiliation{YaLong River Hydropower Development Company, Chengdu 610051}
\author{M.G.~Zhao}
\affiliation{School of Physics, Nankai University, Tianjin 300071}
\author{Z.Y.~Zhou}
\affiliation{Department of Nuclear Physics, China Institute of Atomic Energy, Beijing 102413}
\author{J.J.~Zhu}
\affiliation{College of Physical Science and Technology, Sichuan University, Chengdu 610064}
\author{W.B.~Zhu}
\affiliation{NUCTECH Company, Beijing 10084}
\author{X.Z.~Zhu}
\affiliation{Key Laboratory of Particle and Radiation Imaging (Ministry of Education) and Department of Engineering Physics, Tsinghua University, Beijing 100084}
\author{Z.H.~Zhu}
\affiliation{YaLong River Hydropower Development Company, Chengdu 610051}

\collaboration{CDEX Collaboration}
\noaffiliation


\date{\today}

\begin{abstract}
We report results of a search for light Dark Matter WIMPs with CDEX-1 experiment at
the China Jinping Underground Laboratory, based on 53.9 kg-days of data from
a p-type point-contact germanium detector enclosed by a NaI(Tl) crystal
scintillator as anti-Compton detector.
The event rate and spectrum above the analysis threshold of 475 eVee are consistent
with the understood background model.
Part of the allowed regions for WIMP-nucleus coherent elastic scattering at WIMP mass
of 6-20 GeV are probed and excluded.
Independent of interaction channels, this result contradicts the interpretation that
the anomalous excesses of the CoGeNT experiment are induced by Dark Matter,  since
identical detector techniques are used in both experiments.

\end{abstract}


\maketitle

Current direct-detection dark matter experiments aim at searching of the Weakly Interacting Massive Particles
(WIMPs, denoted by $\chi$) via elastic scattering
of nuclei in terrestrial detectors: $\chi+N\rightarrow\chi+N$~\cite{cdmpdg12}. Of particular interest are the potential positive signatures implied by data from the DAMA~\cite{dama}, CoGeNT~\cite{cogent}, CRESST-II~\cite{cresst2} and CDMS-II(Si)~\cite{cdms2si} experiments. Such interpretations, however, are in conflict with the null results from other experiments~\cite{texono13,lux,xenon100,cdmslite,supercdms}. Additional experiments with improved sensitivities to probe this parameter space are crucial.

Germanium detectors with sub-keV sensitivities were identified~~\cite{ulege}
 as effective means to probe the light WIMP regions, motivating development of point-contact germanium detectors ($\ppcge$)~\cite{ppcge} and various experimental searches~\cite{texono09,texono13,cogent,cdex1,malbek}.
In particular, the CoGeNT experiment~\cite{cogent} with a 443 g $\ppcge$ detector reported possible WIMP-induced events as well as annual modulation signatures.

The China Dark Matter Experiment(CDEX) pursues direct searches of light WIMPs towards the goal of a ton-scale germanium detector array at the China Jinping Underground Laboratory(CJPL)~\cite{cjplnews,cdexcosmic} located in Sichuan, China, with about 2400 m of rock overburden. The combined cosmic-ray direct and induced rates at CJPL has been measured~\cite{cdexcosmic} to be 61.7~m$^{-2}$~yr$^{-1}$, consistent with expectations.
Studies from a prototype CDEX-0 detector array with 20~g target mass at CJPL were reported~\cite{cdex0}.
Our earlier measurement at CJPL from the first phase of CDEX experiment (CDEX-1)~\cite{cdex1} is with a $\ppcge$ of target mass 994 g and analysis threshold of 400 eVee (``ee'' denotes electron-equivalent energy), but in the absence of Anti-Compton (AC) detector and prior to surface event suppression. We report new CDEX-1 results in this article with these two crucial features incorporated. A cylindrical NaI(Tl) crystal scintillator with a well-shaped cavity enclosing the $\ppcge$ target serves as the AC-detector. Identification of surface background and derivation of efficiency factors follow the procedures of Ref.~\cite{texono14}. Details of the hardware setup and shielding configurations can be referred to Refs.~\cite{cdex1,cdexshielding}.

\begin{figure}[t] 
{\bf  (a)}
\includegraphics[width=8.5cm,height=5.0cm]{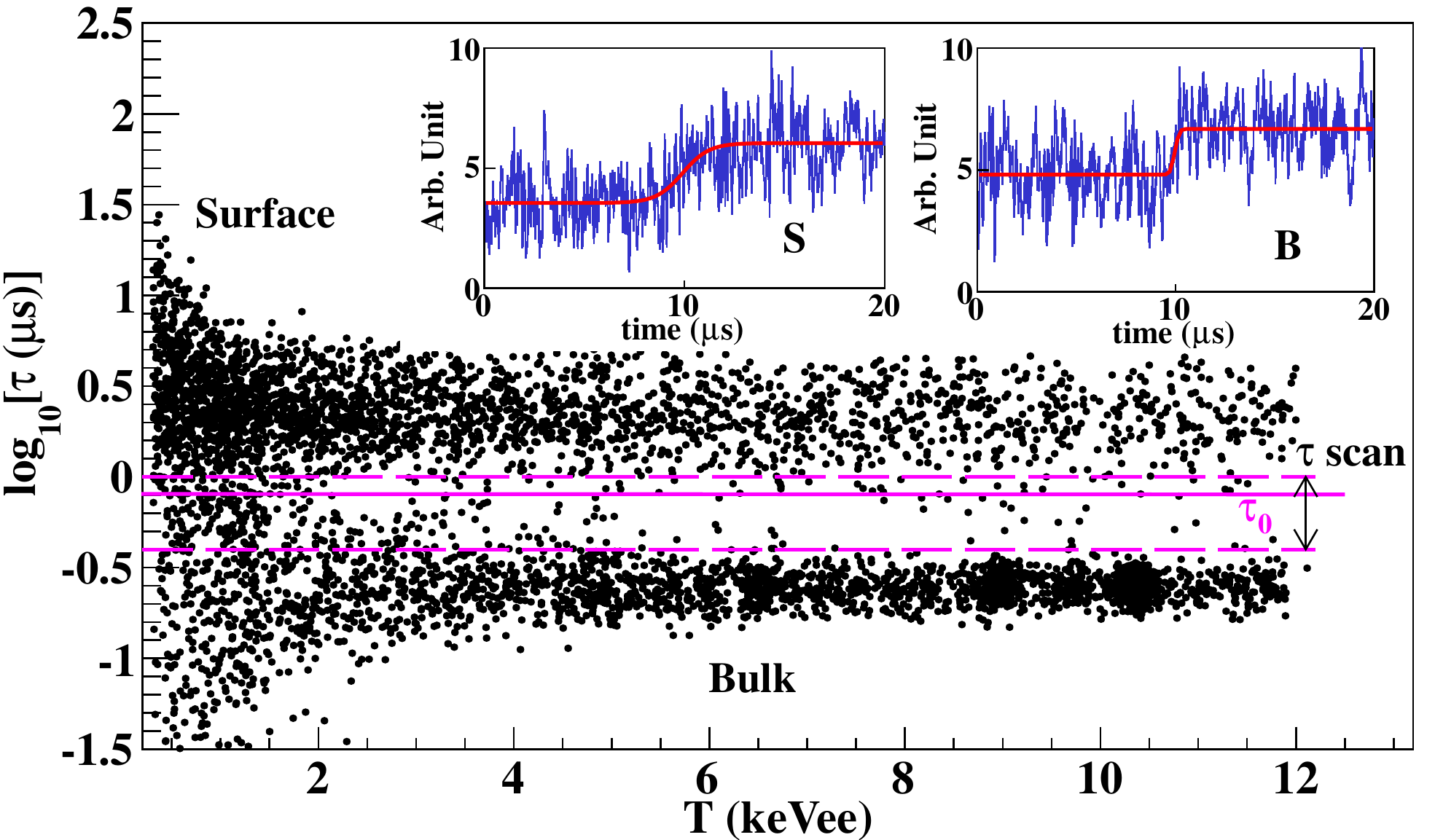}
{\bf (b)}\\
\includegraphics[width=8.5cm,height=3.0cm]{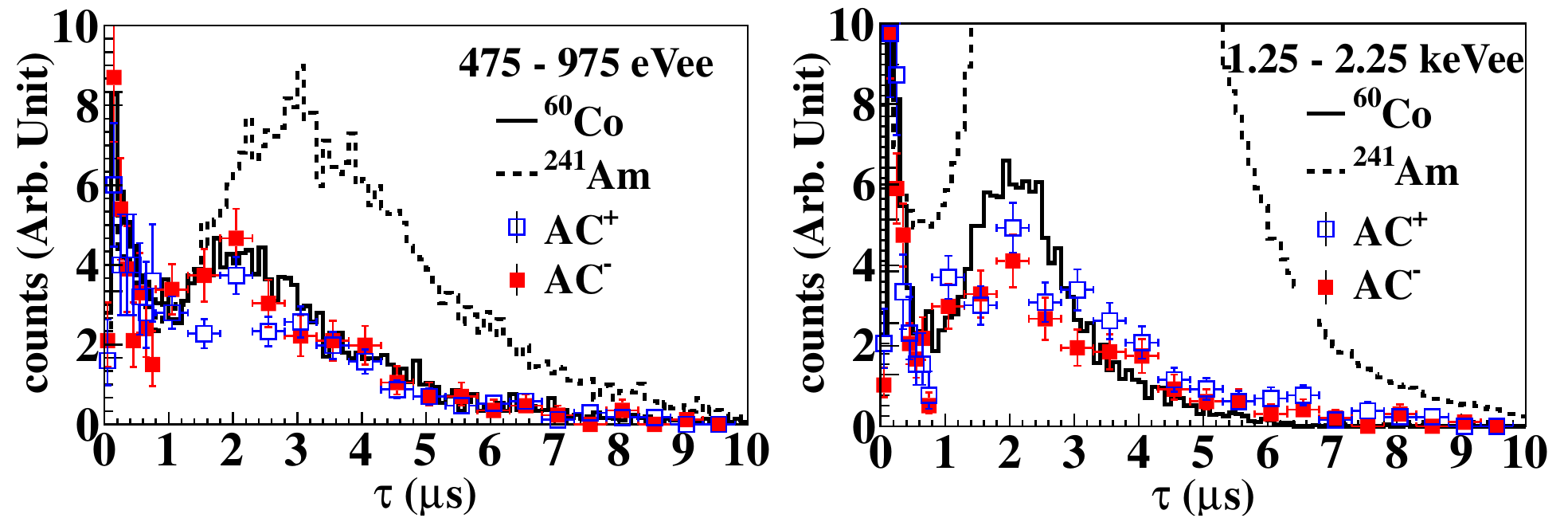}
\caption{
(a) Scatter plot of the $\ppcge$ rise time (${\rm log_{10}}[\tau]$)
versus deposited energy for AC$^{-}$ events. The $\tau_0$-line corresponds to the BS cut in this analysis, with dashed lines indicating the range of cut-stability test. Typical B(S) pulses at energy of $\sim$500~eVee are depicted in the insets. (b) $\tau$-distributions
with $^{241}$Am and $^{60}$Co $\gamma$-sources, together with that from AC$^{+}$ and AC$^{-}$ background data after BS cut,
in the two energy bins of 475-975 eVee (left) and 1.25-2.25 keVee (right).
}
\label{fig::bscut}
\end{figure}

Signals from the $p^{+}$ point-contact electrode are processed by a pulsed reset preamplifier with three identical outputs. Two of them are distributed to the shaping amplifiers at 12~$\mu$s and 6~$\mu$s shaping time which also provides the trigger for data acquisition (DAQ), as well as the energy measurement (T). The remaining one is loaded to a timing amplifier (TA) which provides the fast rise-time information. The outputs are digitized by flash analog-to-digital converters at 100~MHz. A total of 58.7 days of data  was recorded at a trigger rate of $\sim$5~Hz. The DAQ dead time is 0.1\%, as measured by events due to random triggers (RT) generated by a precision pulser. Energy calibration was achieved by the cosmogenic X-ray peaks and the zero-energy was defined by the pedestals of RT events. Deviations from linearity is less than 0.8\%.
The trigger efficiency was unity above 320 eVee, as verified by {\it in situ} physics events via an extrapolation of the amplitude distributions to sub-noise edge energy~\cite{texono09}.
The selection of candidate events based on timing correlation and basic pulse shape discrimination
(``Basic Cut'', denoted by BC) as well as the derivation of their efficiencies were discussed
in our earlier report~\cite{cdex1}. The microphonics effects and electronic events induced
by the preamplifier reset timing were completely suppressed and the combined efficiencies
of 86.3\% were accurately evaluated. Events in anti-coincidence (coincidence) with the
AC-detector are denoted by AC$^-$($^+$). The AC$^-$ selection discriminate $\gamma$-ray
induced background at a signal efficiency of $\sim$100\%, as measured by RT events.


\begin{figure}[t] 
{\bf (a)}\\
\includegraphics[width=8.5cm,height=2.8cm]{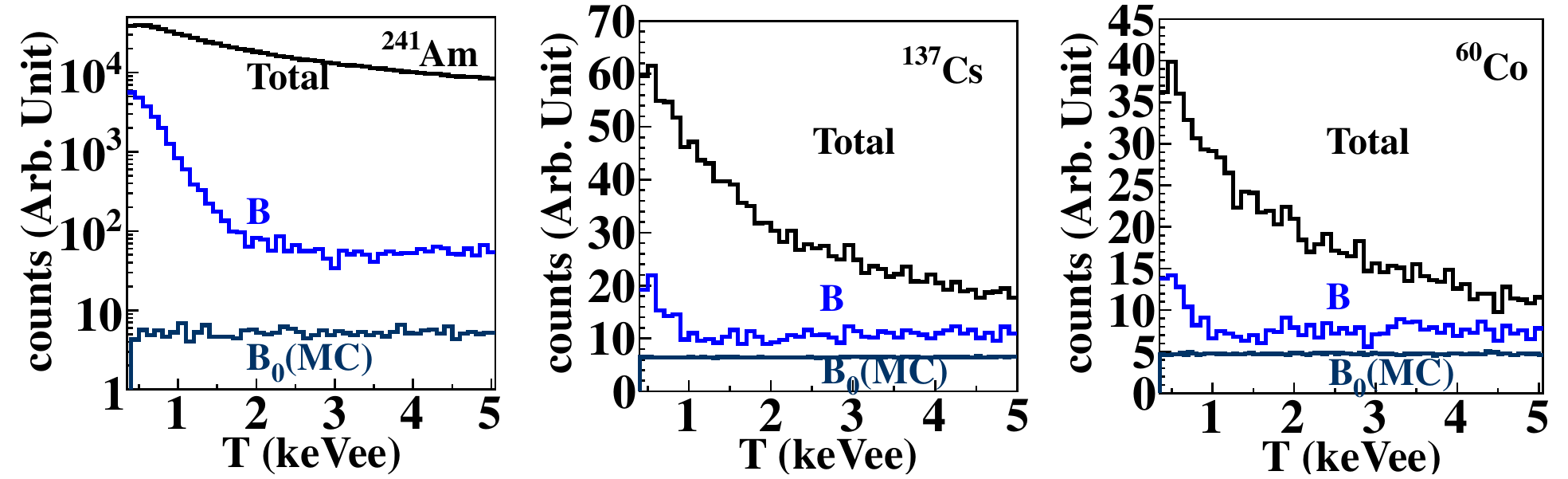}
{\bf (b)}\\
\includegraphics[width=8.5cm,height=3.2cm]{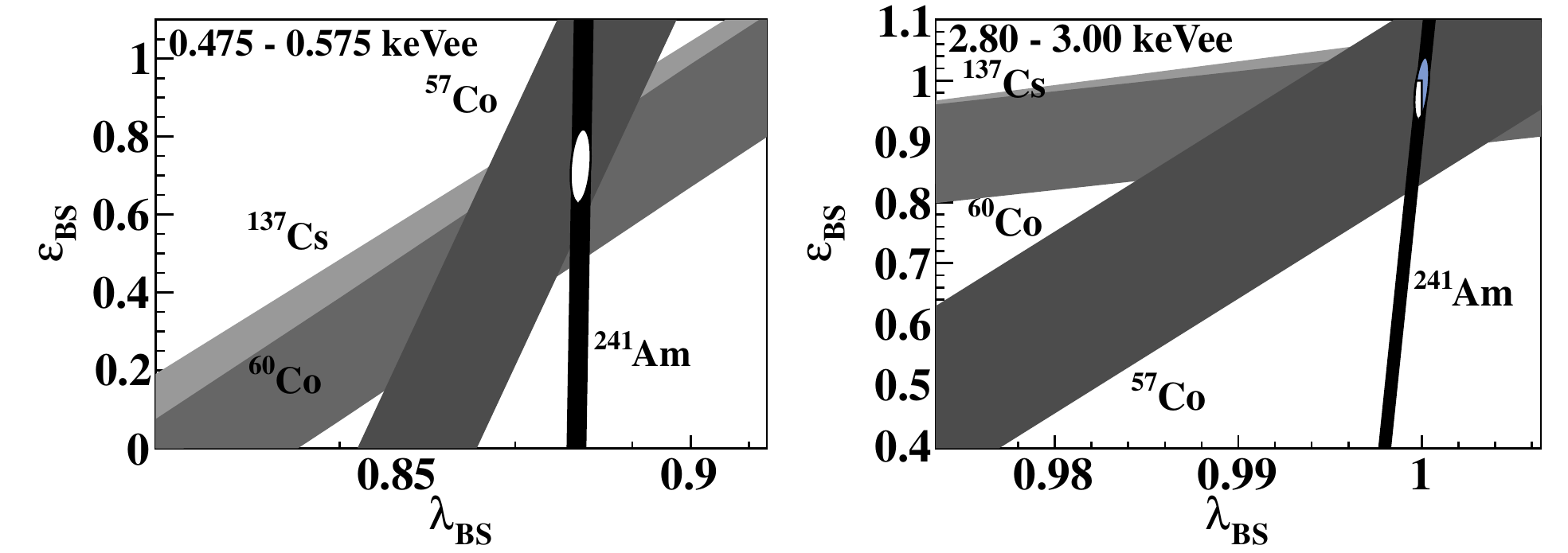}
{\bf (c)}\\
\includegraphics[width=8.5cm,height=2.7cm]{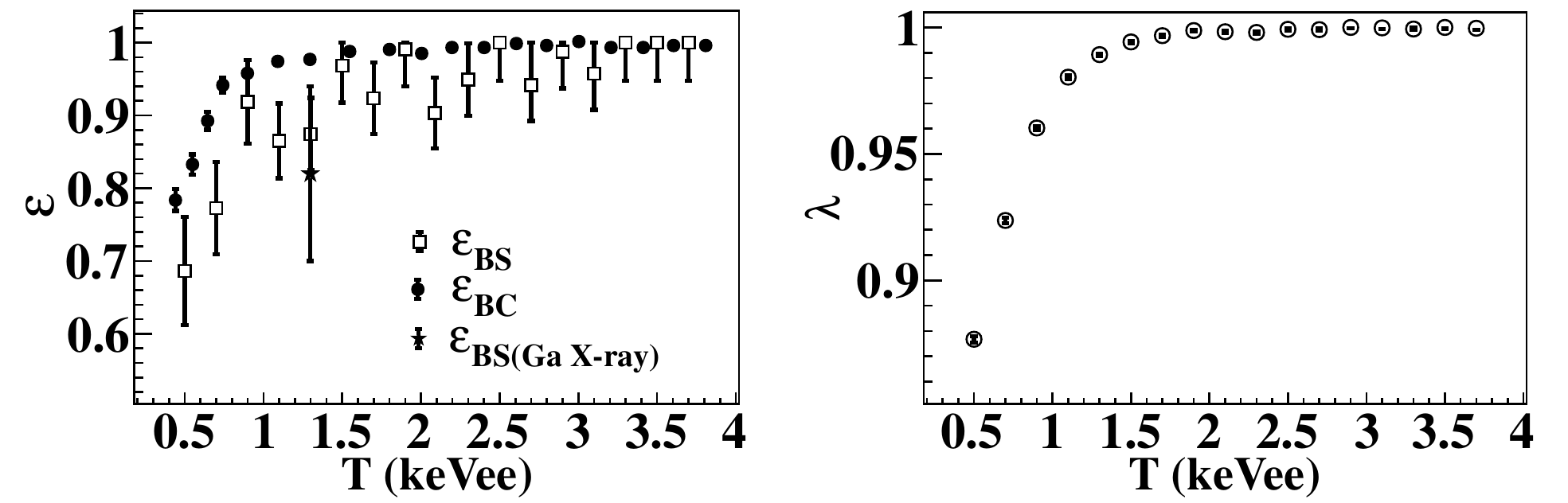}
\caption{
Derivation of $( \effbs , \lmbdbs )$ with $^{241}$Am, $^{60}$Co, $^{137}$Cs and
$^{60}$Co sources. (a) The measured Total and B spectra, in comparison with $B_0$ from simulations.
(b) Allowed bands at 475-575 eVee and at 2.8-3.0 keVee.
(c) The measured ($\effbs$,$\lmbdbs$)
and $\varepsilon_{\rm BC}$ as functions of energy.
Independent measurement on $\effbs$ with
Ga-L X-rays is included.
}
\label{fig::bscalib}
\end{figure}

\renewcommand\tablename{TABLE}
\renewcommand{\thetable}{\arabic{table}}
\begin{table*}
\begin{ruledtabular}
\caption{The various contributions to the total error of AC$^-$$\otimes$B$_0$
at threshold and at a typical high energy bin.}
\centering
\begin{tabular}{lcc}
Energy Bin & 0.475-0.575 keVee & 1.975-2.075 keVee \\
AC$^-$$\otimes$B$_0$ and Errors (kg$^{-1}$keVee$^{-1}$day$^{-1}$) & $ 4.09 \pm 1.47 [stat.] \pm 0.87 [sys.]$ & $4.22 \pm 0.97 [stat.] \pm 0.27 [sys.]$\\
& $=4.09 \pm 1.71 $ & $=4.22 \pm 1.01 $ \\ \hline
I) Statistical Uncertainties (Combined) : & 1.47 & 0.97 \\
    \hspace{0.5cm}(i)Uncertainties on Calibration ($\varepsilon_{BS}$,$\lambda_{BS}$) : & \hspace{1cm}0.32 & \hspace{1cm}0.08 \\
    \hspace{0.5cm}(ii)Derivation of ($\varepsilon_{BS}$,$\lambda_{BS}$)-corrected Bulk Rates : & \hspace{1cm}1.43 & \hspace{1cm}0.97 \\
II) Systematic Uncertainties (Combined) : & 0.87 & 0.27 \\
    \hspace{0.5cm}(i) Rise-time Cut-Value $\tau_{0}$ & \hspace{1cm}0.27 & \hspace{1cm}0.12 \\
    \hspace{0.5cm}(ii) Fiducial Mass & \hspace{1cm}0.05 & \hspace{1cm}0.05 \\
    \hspace{0.5cm}(iii) Normalization Range (3-5 keVee) & \hspace{1cm}0.07 & \hspace{1cm}0.01 \\
    \hspace{0.5cm}(iv) ($B_{0}$,$S_{0}$) = (B,S) at Normalization & \hspace{1.0cm}0.10 & \hspace{1cm}0.10 \\
    \hspace{0.5cm}(v) Choice of Discard Region & \hspace{1cm}0.30 & \hspace{1cm}0.06 \\
    \hspace{0.5cm}(vi) Source Location & \hspace{1cm}0.28 & \hspace{1cm}0.19 \\
    \hspace{0.5cm}(vii) Source Energy Range and Spectra & \hspace{1cm}0.72 & \hspace{1cm}0.12 \\
\end{tabular}
\end{ruledtabular}
\end{table*}

The $n^+$ surface electrode of $\ppcge$ is fabricated by diffusion of lithium ions, having a finite thickness
and producing events with anomalous charge collection and rise time (${\rm \tau}$) ~\cite{surface,texono14}.
The thickness of the Surface layer, including both the dead and inactive layers,
was derived to be (0.97$\pm$0.15)~mm, via the comparison of measured and
simulated intensity ratios of various $\gamma$-peaks from a $^{133}$Ba source, with
the thickness of the copper cryostat being the leading contribution to the uncertainties~\cite{ppcmj12}.
This gives rise to a fiducial mass of 919~g and data strength of 53.9 kg-days for this analysis.
Surface events have anomalous charge collection and risetime ($\tau$) distributions.
The $\tau$ values are measured by fitting to a hyperbolic tangent function to
the TA signals~\cite{texono14}. The $\tau$-distribution of AC$^{-}$ events is depicted in Figure~\ref{fig::bscut}a, showing the two-band structure characterizing bulk (B) and surface (S) events. Typical B and S events
at 500 eVee, together with their fitted-profiles, are displayed in the inset. The $\tau$ distributions
of AC$^-$ and AC$^+$ events at two energy-bands are depicted in of Figure~\ref{fig::bscut}b,
together with those from $^{60}$Co $\gamma$-source. The three samples match well, indicating that
the {\it in situ} background are dominated by ambient high energy (MeV range) $\gamma$'s. Events due to
low energy $\gamma$'s from $^{241}$Am $\gamma$-source, also superimposed, show differences
in the $\tau$-distributions in S.

The $\tau$-cut($\tau_0$) for differentiating the observed B and S events is set at 0.8~$\mu$s.
Two factors are necessary to translate the measured rates (B,S) to the actual rates
(B$_0$,S$_0$)- the B-signal retaining ($\effbs$) and S-background rejection ($\lmbdbs$)
efficiencies. These are related by the coupled equations:
\begin{eqnarray}
{\rm B} & = &  \effbs \cdot {\rm B}_0 ~  +  ~ ( 1 - \lmbdbs ) \cdot {\rm S}_0 \\
{\rm S} & = &  ( 1 - \effbs) \cdot {\rm B}_0 ~  +  ~ \lmbdbs  \cdot {\rm S}_0 ~~ . \nonumber
\label{eq::elcoupled}
\end{eqnarray}
The normalization of (B$_0$,S$_0$)=(B,S) is set at the high energy range of 3$-$5 keVee where the separation of the two bands is larger than the $\tau$-measurement resolution.

Calibration data with $^{241}$Am, $^{57}$Co, $^{137}$Cs and $^{60}$Co are adopted to evaluate ($\effbs$,$\lmbdbs$). Figures~\ref{fig::bscalib}a shows the measured B spectra and their corresponding reference B$_0$ derived from simulation. The allowed bands in ($\effbs$,$\lmbdbs$) derived from the calibration data at 475-575~eVee and 2.8-3.0~keVee are illustrated in Figure~\ref{fig::bscalib}b. The bands overlap at a common region indicating the results are valid over the entire energy range of interest. The measured ($\effbs$,$\lmbdbs$) as a function of energy are depicted in Figure~\ref{fig::bscalib}c. An additional consistency measurement is provided by the ratio of the Ga-L X-rays at 1.3 keVee after $\tau$ selection to its original intensity predicted by the Ga-K X-ray at 10.37~keVee.

\def\cpkkd{\rm{kg^{-1}keV^{-1}day^{-1}}}
\def\effbs{\epsilon _{\rm BS}}
\def\lmbdbs{\lambda _{\rm BS}}
\def\bzero{B_{\rm 0}}

\begin{table*}
\begin{ruledtabular}
\caption{Uncertainty values of ($\effbs$,$\lmbdbs$) and B$_0$
at the most important threshold energy bin.
While systematic errors are important
to the ($\effbs$,$\lmbdbs$) measurements,
the dominant contributions to the uncertainties in the
physics rate B$_0$ remain those from statistical errors
due to limited counts.
Discarding the statistically-strong
$^{241}$Am measurements in a Stress-Test would not introduce
significant changes to B$_0$, on which subsequent physics
analysis are based.}
\begin{tabular}{lccc}
Threshold Bin &
$\effbs ~~ [ \pm \Delta ]$  &
$\lmbdbs ~~ [ \pm \Delta  ]$ &
$\bzero ~~ [ \pm \Delta  ]$ \\
~~ (475-575)~eVee & & & ($\cpkkd$) \\ \hline \hline
Reference Analysis &
$0.757 \pm 0.051(6.7\%)$ [stat.] &
$0.882 \pm 0.001(0.1\%)$ [stat.] &
$4.09 \pm 1.47(35.9\%)$ [stat.] \\
&
\hspace*{4ex} $\pm 0.112(14.8\%)$ [sys.] &
\hspace*{4ex} $\pm 0.020(2.0\%)$ [sys.]&
\hspace*{4ex} $\pm 0.87(21.3\%)$ [sys.] \\ \cline{2-4}
& $= 0.757 \pm 0.123(16.2\%)$ [total] &
$= 0.882 \pm 0.020(2.0\%)$ [total] &
$= 4.09 \pm 1.71(41.8\%)$ [total] \\ \hline
Stress-Test &
$0.673 \pm 0.074(11.0\%)$ [stat.] &
$0.862 \pm 0.007(0.8\%)$ [stat.] &
$3.92 \pm 1.53(39.0\%)$ [stat.] \\
~(Discard $^{241}$Am Data) &
\hspace*{4ex} $\pm 0.112(16.6\%)$ [sys.] &
\hspace*{4ex} $\pm 0.020(2.3\%)$ [sys.]&
\hspace*{4ex} $\pm 0.91(23.2\%)$ [sys.] \\ \cline{2-4}
& $= 0.673 \pm 0.134(20.0\%)$ [total] &
$= 0.862 \pm 0.021(2.4\%)$ [total] &
$= 3.92 \pm 1.78(45.4\%)$ [total] \\ 
\end{tabular}
\end{ruledtabular}
\end{table*}

\begin{figure}[t] 
{\bf (a)}\\
\includegraphics[width=8.5cm,height=4.0cm]{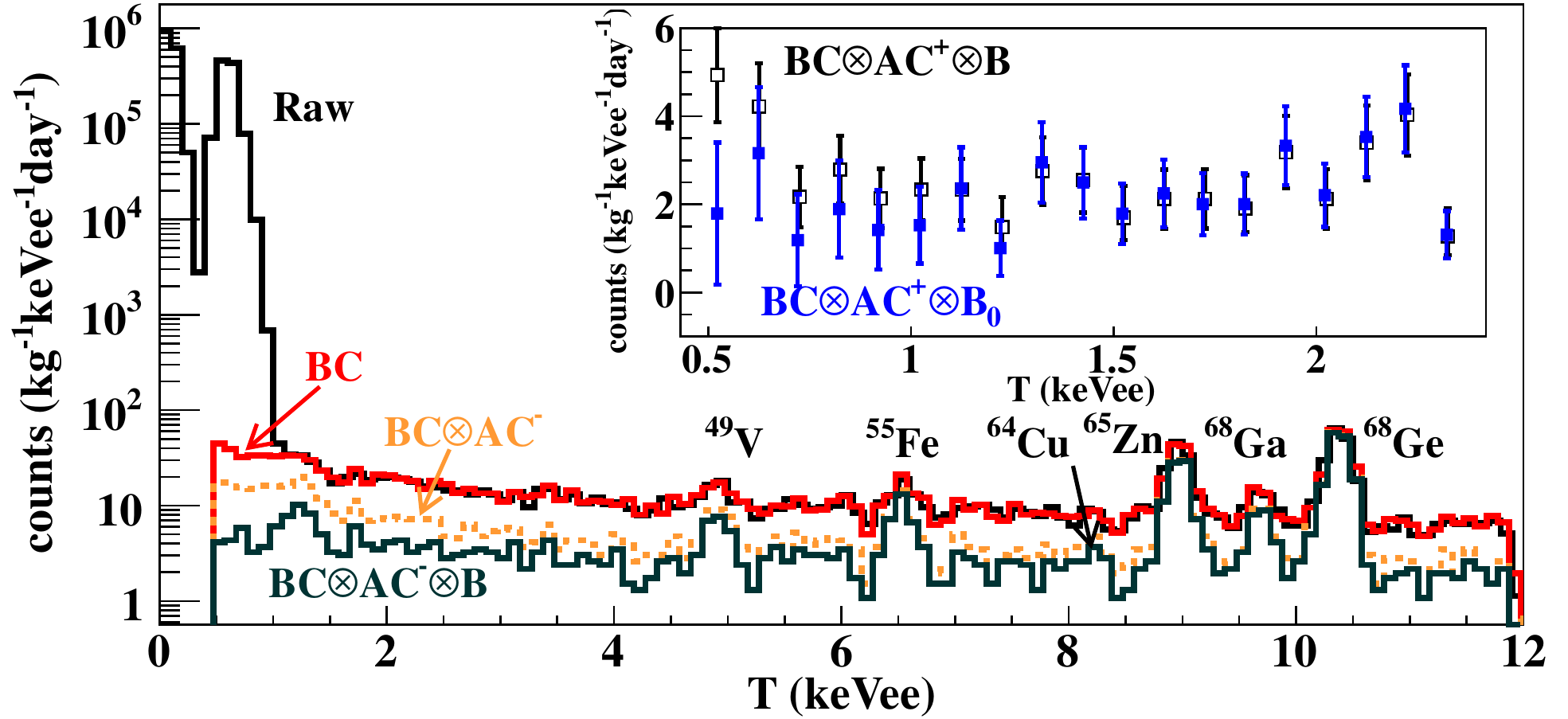}\\
{\bf (b)}\\
\includegraphics[width=8.5cm,height=3.5cm]{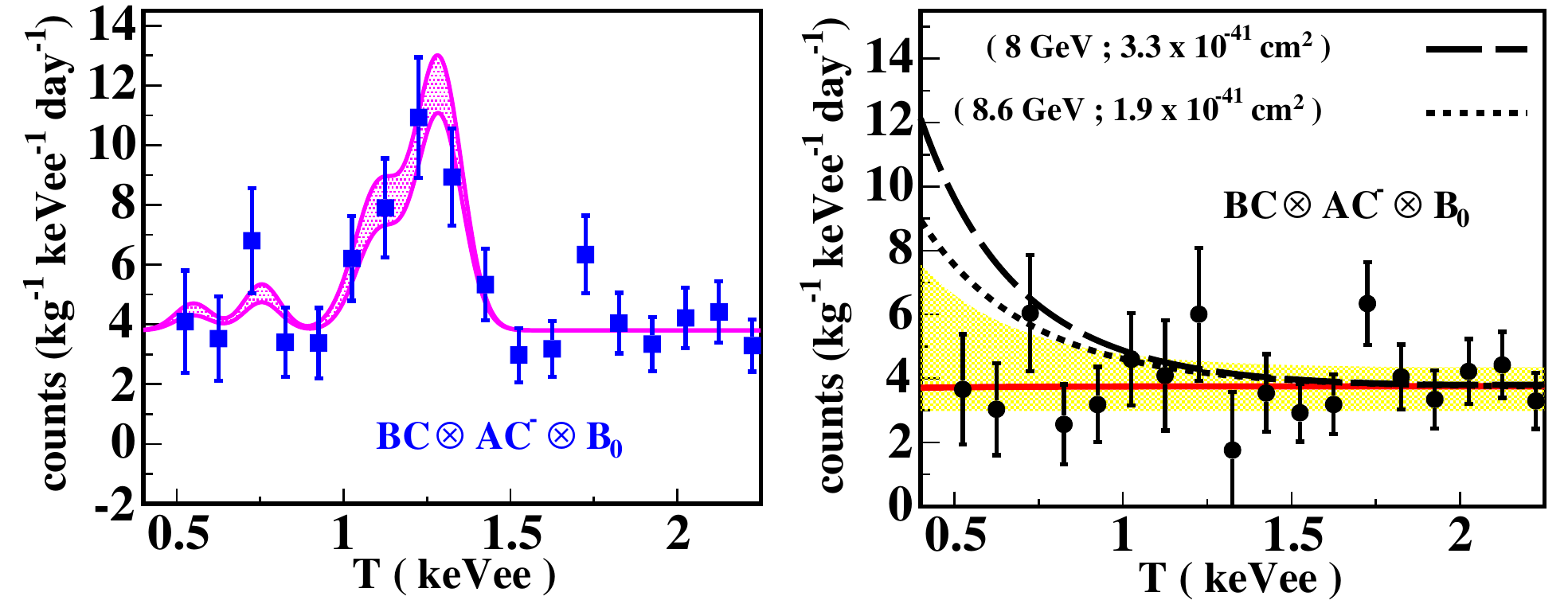}\\
\caption{
(a)Measured energy spectra, showing the
raw data with those at various stages of the selection procedures including the BC and AC and BS cuts.
The inset shows the B spectra of AC$^{+}$, before and after ($\effbs$,$\lmbdbs$) correction.
(b)
The ($\effbs$,$\lmbdbs$)-corrected BC$\otimes$AC$^-$$\otimes$B$_0$ spectrum~-~
Left: the L-X lines predicted by the K-X intensities; Right: Residual with the L-X peaks subtracted, together with the best fit profile and 2$\sigma$ uncertainty band at $\mwimp$=~8~GeV, as well as recoil spectra at the best-fit values of CoGeNT-2013~\cite{cogent} and CDMS(Si)~\cite{cdms2si} at ($\mwimp$~;~$\csnospin$) = (8~GeV~;~3.3~x~10$^{-41}$~cm$^2$) and (8.6~GeV ; 1.9~x~10$^{-41}$~cm$^2$), respectively.
}
\label{fig::spectra}
\end{figure}

The raw spectrum and those at various stages of selection procedures are depicted in Figure~\ref{fig::spectra}a.
The peak at 600~eVee is due to induced electronic noise from preamplifier resets and is completely rejected by timing correlation in BC~\cite{cdex1}. The ($\effbs$,$\lmbdbs$)-corrected spectra of the candidate events, defined as AC$^{-}$$\otimes$B$_0$ and shown in the Figure~\ref{fig::spectra}b, can be derived via the solution of Eq. 1:

\begin{eqnarray}
{\rm B}_0 & = &  \frac{\lmbdbs}{\effbs + \lmbdbs - 1} \cdot {\rm B} ~  +  ~\frac{\lmbdbs - 1 }{\effbs + \lmbdbs - 1 }   \cdot {\rm S} ~~\\
{\rm S}_0 & = &  \frac{\effbs - 1}{\effbs + \lmbdbs - 1}\cdot {\rm B} ~  + ~\frac{\effbs }{\effbs + \lmbdbs - 1}\cdot {\rm S} ~~ . \nonumber
\label{eq::elsol}
\end{eqnarray}

The peaks correspond to known K-shell X rays from the cosmogenically-activated isotopes.
The analysis threshold is placed at 475~eVee, below which the sensitivity is constrained by the noise edge. The AC$^+$ spectra, depicted in the inset of Figure~\ref{fig::spectra}a,
correspond to events from ambient $\gamma$-rays which are in coincidence with the NaI(Tl) detector.
An expected flat AC$^+$$\otimes$B$_0$ spectrum down to threshold is obtained, demonstrating that the ($\effbs$,$\lmbdbs$)-correction is valid.

The various components which contribute to the errors of AC$^-$$\otimes$B$_0$
at threshold and at a typical high energy bin are summarized in Table 1.
Uncertainty values of ($\effbs$,$\lmbdbs$) and B$_0$ at threshold
are listed in Table 2.
Systematic uncertainties originate from (i) parameter choices of the
analysis procedures and (ii) possible differences in the locations
and energy spectra between the calibration sources and background events.
The combined errors of AC$^-$$\otimes$B$_0$ are dominated by the statistical
uncertainties in the measurement of (B,S), boosted by the factor of [1/($\effbs$+$\lmbdbs$-1)]
from Eq. 2 as ($\effbs$,$\lmbdbs$) deviates from unity below 1.5~keVee.
Contributions of systematic uncertainties are minor (increasing the total error of AC$^-$$\otimes$B$_0$
from 1.47 to 1.71 kg$^{-1}$keVee$^{-1}$day$^{-1}$ at threshold)
and are taken into account in the analysis.
The variations of the key parameters over changes of
$\tau_0$ within the $\tau$-scan range in Figure~\ref{fig::bscut}a are studied.
The B$_0$ spectra are stable, robust and independent of
$\tau_0$, as indicated by the small variations relative to the uncertainties.

The high statistics of $^{241}$Am surface events produce the narrow vertical bands in Figures~\ref{fig::bscalib}b, which in turn drives the small statistical error of $\lmbdbs$ in Figures~\ref{fig::bscalib}c. An additional stress-test was performed. The $^{241}$Am measurements are discarded altogether, and ($\effbs$,$\lmbdbs$) are derived with the $^{57}$Co, $^{137}$Cs and $^{60}$Co data. The results are summarized in Table 2.
The B$_0$ at the threshold energy bin would be shifted only by 4.2\% from 4.09 to 3.92 kg$^{-1}$keVee$^{-1}$day$^{-1}$. The effects on the subsequent physics analysis are therefore negligible.

\begin{figure}[t] 
\includegraphics[height=7.3cm,width=8.5cm]{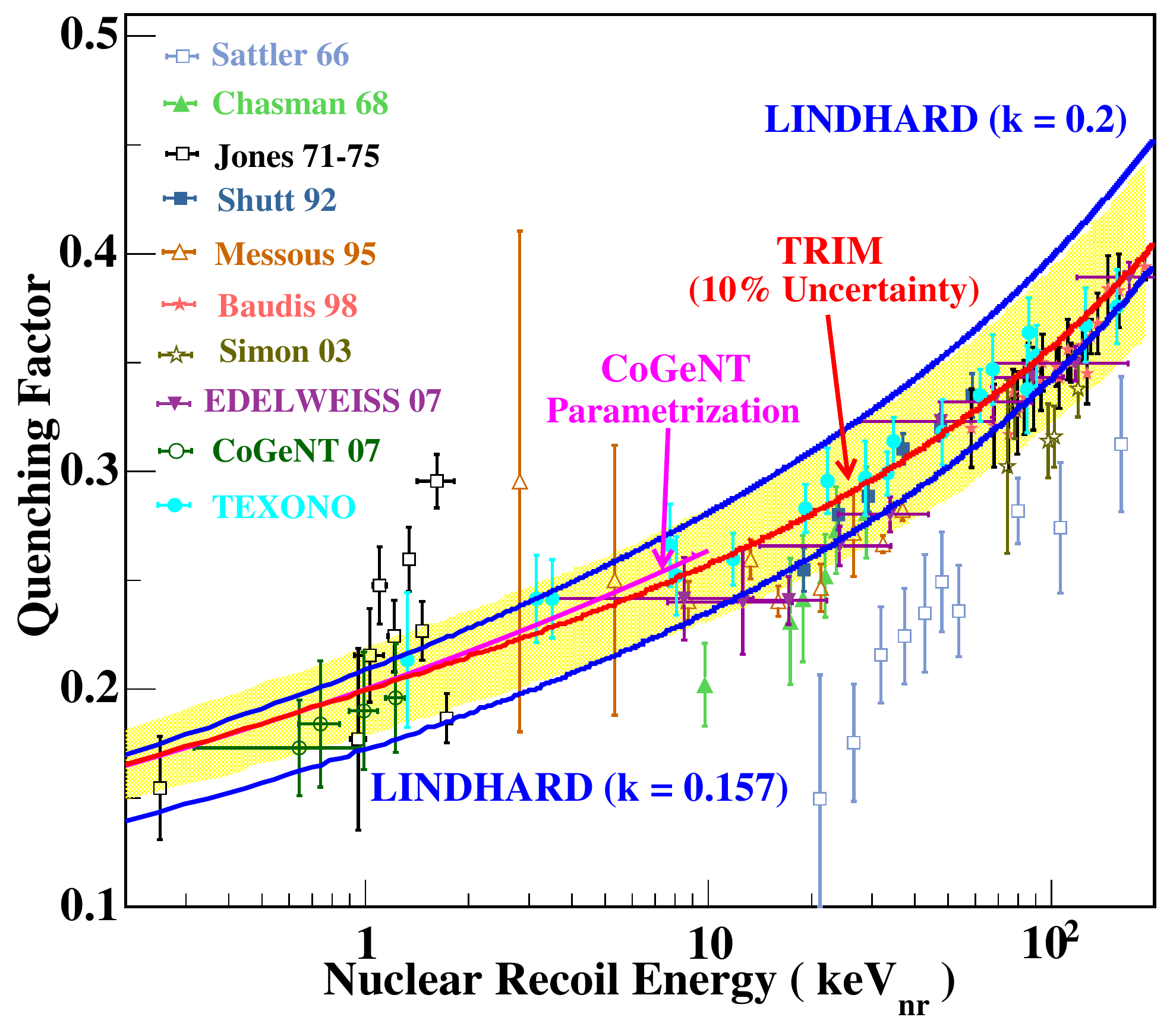}
\caption{ The QF curve which is derived from TRIM~\cite{trim} as a function of nuclear recoil energy, together with a $\pm$10\% band (light shadow band). The various experimental measurements are overlaid, so are the alternative QFs from parametrization of CoGeNT~\cite{cogent} and the Lindhard theory~\cite{lindhard} at k=0.2 and k=0.157 adopted by CDMSlite~\cite{cdmslite}. It can be see that the TRIM results with uncertainties covers most data points as well as the alternative formulations.
}
\label{fig::qf}
\end{figure}

High energy $\gamma$-rays from ambient radioactivity produce
flat electron-recoil background at low energy, as predicted
by simulations and is verified by the $^{241}$Am, $^{137}$Cs and
$^{60}$Co spectra of Figure~\ref{fig::bscalib}a. The L-shell
X-ray lines are predicted by the K-shell peaks. Both
background are subtracted from the $(\effbs,\lmbdbs)$-corrected AC$^-$$\otimes$B$_0$ spectrum
as shown in Figure~\ref{fig::spectra}b. A minimum-$\chi^2$ analysis is applied to the
residual spectrum within 0.475 and 2.25~keVee, adopting two free and positive definite
parameters which characterize the flat ambient
$\gamma$-background and the possible $\chi$-N spin-independent cross-section ($\csnospin$),
respectively. Conventional astrophysical models~\cite{cdmpdg12} are adopted to
describe WIMP-induced interactions, using the local WIMP density of 0.3~GeV/cm$^3$,
 the Maxwellian velocity distribution with
$v_0$=220~km/s and the galactic escape velocity of $v_{esc}$=544~km/s.
The quenching function (QF) in Ge is evaluated with the TRIM software package~\cite{trim}. The derived QF is depicted in Figure~\ref{fig::qf} with measured data~\cite{geqf} showing good agreement over a large range of nuclear recoil energy.
A systematic uncertainty of 10\% is taken, corresponding to the spread of
individual data points, as well as the deviations with the alternative Lindhard model~\cite{lindhard}.
Analysis is performed by scanning QF within $\pm$10\% of their nominal value,
and the most conservative constraints are adopted as the limits.

\begin{figure}[t] 
\includegraphics[height=8.0cm,width=8.6cm]{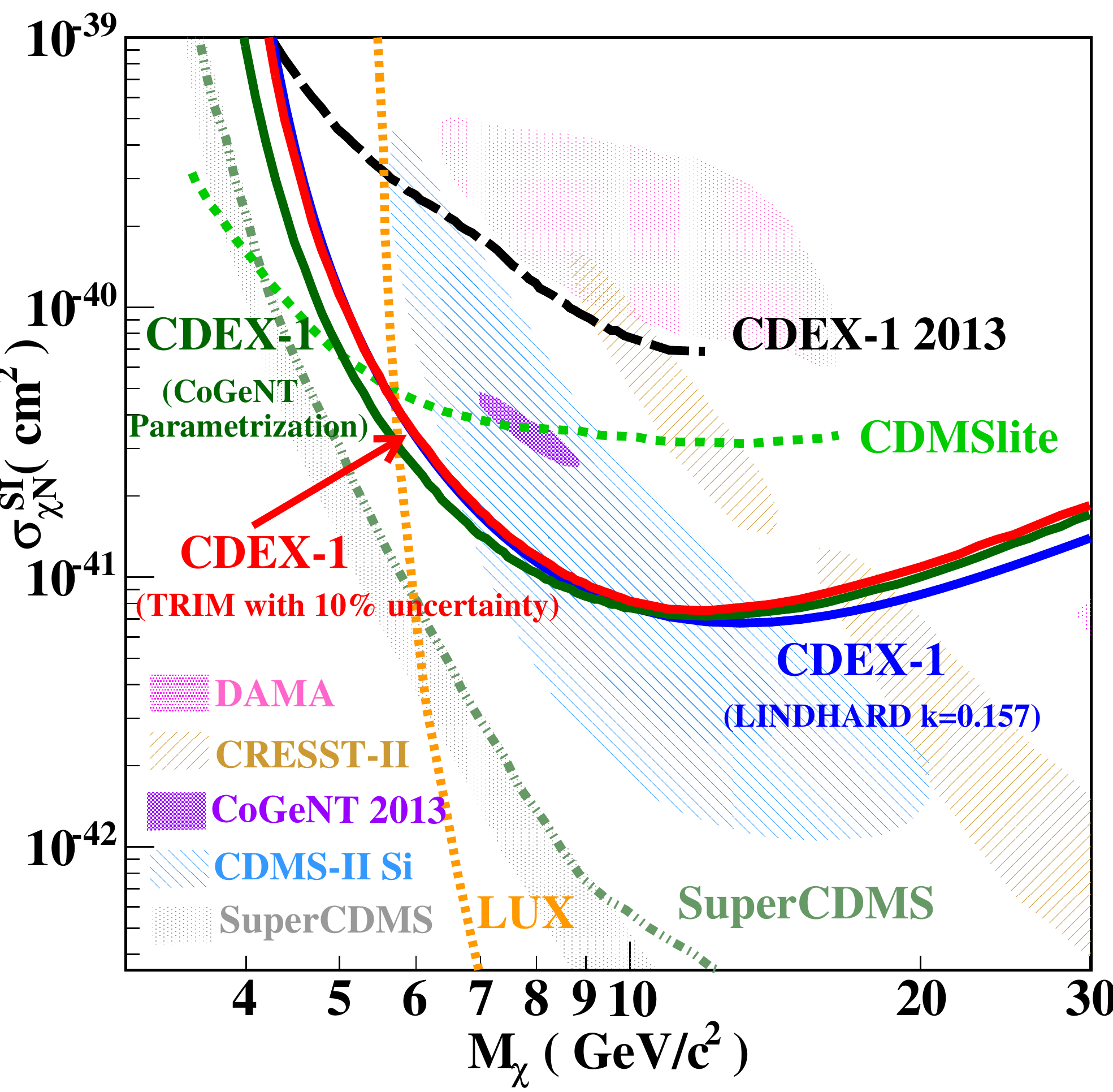}
\caption{
The 90\% confidence level upper limit of spin-independent $\chi$N coupling.
The CDEX-1 results from this work is depicted in solid red. Bounds from alternative
QF formulations are represented as solid blue and green curves. Results from other benchmark experiments~\cite{dama,cresst2,cogent,cdms2si,supercdms,cdex1,lux,cdmslite}
are superimposed.
}
\label{fig::explot}
\end{figure}

As illustration, the best-fit spectrum at $\mwimp$=8~GeV, where
$\csnospin$=(-1.80$\pm$9.28)x10$^{-42}$~cm$^2$ at $\chi^2$/dof=8.11/16
(p-value$=$0.95), is depicted in Figure~\ref{fig::spectra}b, with the band
representing the 2$\sigma$ uncertainties. Exclusion plot of $\csnospin$ versus $\mwimp$ at 90\% confidence level
is displayed in Figure~\ref{fig::explot}. The bounds from other benchmark
experiments are superimposed~\cite{cogent,cdmslite,lux,supercdms}.
As comparison, different QFs (parametrization of CoGeNT~\cite{cogent} and Lindhard theory with k=0.157 of CDMSlite~\cite{cdmslite}) are used to derive alternative exclusion curves, also displayed in Figure~\ref{fig::explot}. It can be seen that the analysis procedures adopted in this work provide the most conservative constraints.

An order of magnitude improvement in the sensitivities of $\csnospin$ has been achieved over our previous results~\cite{cdex1}. Part of the light WIMP regions within 6 and 20 GeV implied by earlier experiments are probed and rejected. In particular, the CoGeNT anomalous events at sub-keV energy~\cite{cogent} are not reproduced
in these results based on identical detector techniques. This strongly
disfavors the excess is induced by dark matter, independent of interaction channels.
For instance, electromagnetic final states are not constrained by experiments like SuperCDMS~\cite{supercdms} which measure nuclear recoil events. They can, however, be probed by the CDEX-1 data.

The CDEX-1 experiment continues to accumulate data at CJPL. Research programs are pursued to further reduce the physics threshold via hardware and software efforts. Time modulation of the data will be studied. A PCGe array of 10 kg target mass range enclosed in an active liquid argon anti-Compton detector is being constructed. Feasibility studies towards scale-up to ton-scale experiment~\cite{cdexshielding} are being pursued.

This work was supported by the National Natural Science Foundation of China
(contract numbers: 10935005, 10945002, 11275107, 11175099) and National Basic
Research program of China (973 Program) (contract number: 2010CB833006) and NSC
99-2112-M-001-017-MY3 and Academia Sinica Principle Investigator Award 2011-
2015 from Taiwan.

\end{document}